\documentclass{article}
\usepackage{spconf,amsmath}
\usepackage{graphicx}
\usepackage{amsmath,amsfonts}
\usepackage{bm}
\usepackage[legacycolonsymbols]{mathtools}

\def\mysection#1{\section{#1}}
\def\mysubsection#1{\subsection{#1}}

\allowdisplaybreaks
\def\NamesSpace{\quad\!\!}

\title{Real-time speech extraction using\\ spatially regularized independent low-rank matrix analysis and \\rank-constrained spatial covariance matrix estimation}

\name{Yuto Ishikawa$^{\star}$ \NamesSpace Kohei Konaka$^{\star}$ \NamesSpace Tomohiko Nakamura$^{\dagger}$ \NamesSpace Norihiro Takamune$^{\star}$ \NamesSpace Hiroshi Saruwatari$^{\star}$
\thanks{
  This work was supported by JST Moonshot R\&D Grant Number JPMJMS2011 (for algorithm development) and Tateisi Science and Technology Foundation (for practical experiment).
}}
\address{$^{\star}$ The University of Tokyo, Graduate School of Information Science and Technology, Tokyo, Japan\\
$^{\dagger}$ The National Institute of Advanced Industrial Science and Technology (AIST), Tokyo, Japan}

\begin{document}
\ninept
\maketitle
\begin{abstract}

Real-time speech extraction is an important challenge with various applications such as speech recognition in a human-like avatar/robot.
In this paper, we propose the real-time extension of a speech extraction method based on independent low-rank matrix analysis~(ILRMA) and rank-constrained spatial covariance matrix estimation~(RCSCME).
The RCSCME-based method is a multichannel blind speech extraction method that demonstrates superior speech extraction performance in diffuse noise environments.
To improve the performance, we introduce spatial regularization into the ILRMA part of the RCSCME-based speech extraction and design two regularizers.
Speech extraction experiments demonstrated that the proposed methods can function in real time and the designed regularizers improve the speech extraction performance.

\end{abstract}
\mysection{Introduction}
\label{sec:intro}

Multichannel blind speech extraction (BSE) aims at extracting the target speech from noisy mixture signals recorded by a microphone array without any prior information \cite{Takahashi2009TASLP,Kubo2020TASLP}.
It is essential for various speech applications.
For example, it can be employed in a human--avatar/robot communication system.
In a practical (i.e., noisy) environment, BSE can enhance smooth communication by reducing background noise, which often decreases speech recognition performance and increases the listening effort of the avatar/robot operator.
Another crucial property of BSE for facilitating smooth communication is its capability to function in real time.
The processing delay of BSE affects all subsequent modules of the system.
In this paper, we address the real-time BSE problem in which the observed mixture signals consist of a single directional speech and diffuse background noise.

A standard approach to developing a real-time BSE method is to extend an offline BSE method to function in real time.
We previously proposed a state-of-the-art offline BSE method based on independent low-rank matrix analysis (ILRMA)~\cite{Kitamura2016TASLP,Kitamura2018ASS} and rank-constrained spatial covariance matrix estimation (RCSCME)~\cite{Kubo2020TASLP}.
ILRMA is one of the state-of-the-art blind source separation (BSS)\cite{Sawada2019APSIPA} methods for a (over-)determined case, i.e., the number of microphones $M$ is greater than or equal to that of sources $N$.
It estimates linear demixing filters from the observed mixture signals on the basis of the statistical independence between sources.
Under diffuse noise, ILRMA cannot separate only the target speech (i.e., the separated speech contains residual diffuse noise), but it can cancel only the target speech accurately~\cite{Araki2003EURASIP}.
RCSCME utilizes the latter property and extracts the target speech from the observed mixture signals by referring to the estimated demixing filters (particularly for diffuse noise).
Although RCSCME mainly consists of scalar operations and is suited for real-time processing, ILRMA requires many matrix operations and is computationally more expensive than RCSCME.
Thus, it is difficult to extend the RCSCME-based method for real-time processing in a straightforward manner.

In this paper, we propose the real-time extension of the RCSCME-based speech extraction method using a parallel computing algorithm (blockwise batch algorithm)~\cite{Mukai2004IEICE,Mori2006EURASIP}.
Our idea for the proposed extension stems from the facts that the ILRMA outputs required in RCSCME are only the demixing filters and we can split the entire process of the offline RCSCME-based method into the ILRMA and RCSCME parts.
By using the blockwise batch algorithm, we can parallelize these two parts: the ILRMA part runs across multiple frames, whereas the RCSCME part runs in a frame-by-frame manner.
Thus, the proposed method can function in real time.

We further extend the proposed real-time method to incorporate the approximate speaker direction, which is often available in human--avatar/robot communication systems.
Since ILRMA is a fully blind method, we must determine which demixing filters obtained by ILRMA contain the target speech. We call this process the channel selection.
Although a method based on the maximum kurtosis criterion~\cite{Fujihara2008IWAENC} functions well in the offline scenario, we experimentally found that it often failed in the channel selection in the real-time scenario, which significantly degraded the speech extraction performance.
To mitigate such failures, we replace the usual ILRMA with the spatially regularized ILRMA~\cite{Mitsui2018ICASSP}.
Since the regularizers proposed in \cite{Mitsui2018ICASSP} are not directly applicable to our case, we design two spatial regularizers using the approximate speaker direction and derive parameter estimation algorithms for the spatially regularized ILRMA.
Speech extraction experiments showed the real-time functioning of the proposed methods and the effectiveness of the designed regularizers.

\mysection{Related Offline methods}
\label{sec:works}
\begin{figure*}[t]
  \begin{center}
    \includegraphics[width=0.89\textwidth]{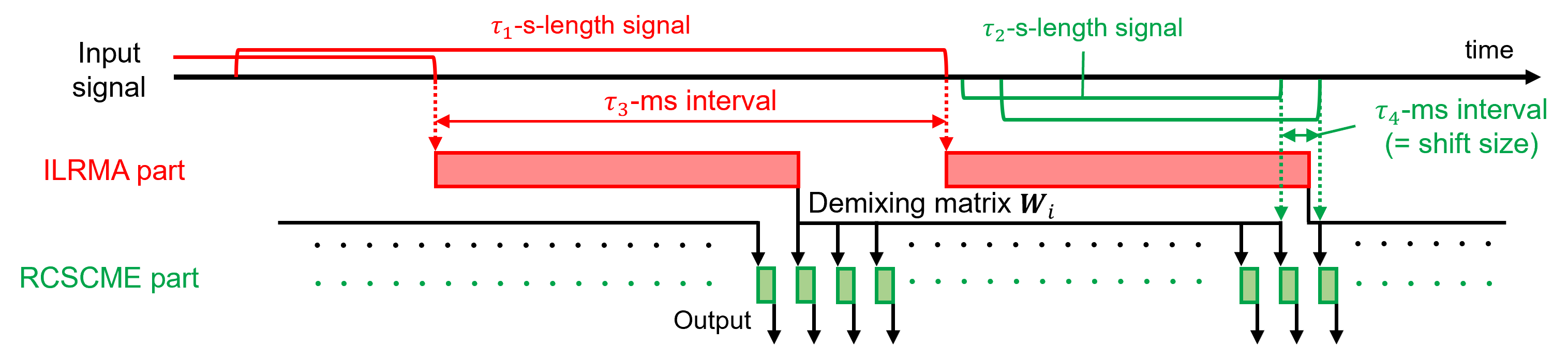}
    \vspace*{-0.5cm}
    \caption{Schematic of parallel processing in real-time RCSCME-based speech extraction method.}
    \vspace*{-0.66cm}
  \end{center}
  \label{fig:framework}
\end{figure*}

\mysubsection{ILRMA~\cite{Kitamura2016TASLP}}
\label{ssec:ILRMA}

Let $\bm{x}_{ij}=(x_{ij1},\cdots,x_{ijM})^{\mathsf{T}}\in\mathbb{C}^{M}$, $ \bm{s}_{ij}=(s_{ij1},\cdots,s_{ijN})^{\mathsf{T}}\in\mathbb{C}^{N}$, and $\bm{y}_{ij}=(y_{ij1},\cdots,y_{ijN})^{\mathsf{T}}\in\mathbb{C}^{N}$ be the short-time Fourier transforms (STFTs) of the observed, source, and separated signals, respectively.
Here, $i\in\{1,\ldots,I\}$, $j\in\{1,\ldots,J\}$, $m\in\{1,\ldots,M\}$, and $n\in\{1,\ldots,N\}$ are the indices of the frequency bins, time frames, microphones, and sources, respectively, and $\cdot^{\mathsf{T}}$ denotes the transpose.
ILRMA assumes that each source is a point source and the reverberation time is sufficiently shorter than the window length of the STFT.
With these assumptions, the observed signal is approximately represented as $\bm{x}_{ij}=\bm{A}_{i}\bm{s}_{ij}$.
Here, $\bm{A}_{i}=(\bm{a}_{i1},\cdots,\bm{a}_{iN})\in\mathbb{C}^{M\times N}$ is the mixing matrix, which represents the time-invariant spatial characteristics of the transmission system, and $\bm{a}_{in}$ is the steering vector of source $n$.
If $M=N$ and $\bm{A}_{i}$ is regular, the separated signals can be obtained as
\begin{align}
  \bm{y}_{ij}=\bm{W}_{i}\bm{x}_{ij},
\end{align}
where $\bm{W}_{i}=(\bm{w}_{i1},\cdots,\bm{w}_{iN})^{\mathsf{H}}=\bm{A}_{i}^{-1}\in\mathbb{C}^{N\times M}$ is the demixing matrix and $\cdot^{\mathsf{H}}$ denotes the Hermitian transpose.

In ILRMA, it is assumed that the separated signal $y_{ijn}$ follows a univariate complex Gaussian distribution whose mean and time-variant variance are zero and $r_{ijn}>0$, respectively.
The variance $r_{ijn}$ is modeled by nonnegative matrix factorization (NMF)~\cite{Lee1999Nature} as $r_{ijn}=\sum_{k}{t_{ikn}v_{kjn}}$, where $t_{ikn}\geq 0$ and $v_{kjn}\geq 0$ are the NMF variables, $k\in\{1,\ldots,K\}$ is the NMF basis index, and $K$ is the number of NMF bases.
The cost function $\mathcal{T}_{\mathrm{ILRMA}}$ is given as the negative log-likelihood of the observed signal:
\begin{align}
  \mathcal{T}_{\mathrm{ILRMA}}
  =&\frac{1}{J}\sum_{i,j,n}{\Biggl(
    \frac{|\bm{w}_{in}^{\mathsf{H}}\bm{x}_{ij}|^{2}}{r_{ijn}}+\log{r_{ijn}}
  \Biggr)}
  \nonumber\\
  &-\sum_{i}{\log{|\det{\bm{W}_{i}}|^{2}}}+\mathrm{const.},
\end{align}
where $\mathrm{const.}$ denotes the term independent of $\bm{w}_{in}$, $t_{ikn}$, and $v_{kjn}$.
The cost function $\mathcal{T}_{\mathrm{ILRMA}}$ is minimized by alternately updating the NMF variables $t_{ikn}$ and $v_{kjn}$, and the demixing matrix $\bm{W}_{i}$.
The NMF variables are updated on the basis of the majorization-minimization algorithm~\cite{Hunter2000JCGS} (see \cite{Kitamura2016TASLP} for the detailed update rules).
The demixing matrix $\bm{W}_{i}$ is updated by iterative projection (IP)~\cite{Ono2011WASPAA}:
\begin{align}
  \label{updateDinNaive}
  \bm{D}_{in} &\leftarrow \frac{1}{J}\sum_{j}\frac{\bm{x}_{ij}\bm{x}_{ij}^{\mathsf{H}}}{\sum_{k}t_{ikn}v_{kjn}},
  \\
  \bm{w}_{in} &\leftarrow (\bm{W}_{i}\bm{D}_{in})^{-1}\bm{e}_{n},
  \\
  \bm{w}_{in} &\leftarrow \frac{\bm{w}_{in}}{\sqrt{\bm{w}_{in}^{\mathsf{H}}\bm{D}_{in}\bm{w}_{in}}},
\end{align}
where $\bm{e}_n$ is the $n$th column vector of the $M\times M$ identity matrix.

\mysubsection{RCSCME~\cite{Kubo2020TASLP}}
\label{ssec:RCSCME}
When we apply ILRMA to mixture signals that consist of a directional target speech and diffuse noise, although $M-1$ separated signals contain only diffuse noise with high accuracy, one separated signal corresponding to the target speech direction includes diffuse noise arriving from that direction~\cite{Takahashi2009TASLP,Araki2003EURASIP}.
From the $M-1$ separated signals that consist of only noise, the rank-($M-1$) component of the spatial covariance matrix (SCM) for noise and the steering vector of the target speech can be estimated with high accuracy.
In RCSCME, the parameters calculated from the output of ILRMA are fixed, and then the deficient rank-1 component of the noise SCM and the time-variant variance of the target speech and noise are estimated.
Finally, a multichannel Wiener filter is applied to the observed signal using the following parameters estimated by ILRMA and RCSCME: the time-variant variance of the target speech and noise, the steering vector of the target speech, the rank-($M-1$) component of the noise SCM, and the complementary rank-1 component and its weights. 
With respect to computational complexity, RCSCME requires a few parameters to be estimated and the most part of RCSCME consists of only scalar operations, which is suitable for real-time processing.

\mysection{Proposed method}
\label{sec:proposal}

\mysubsection{Real-time RCSCME-based speech extraction method}
\label{ssec:realtime}
A simple approach to real-time speech extraction is to finish the entire process within the shift length of the STFT for each frame.
The offline RCSCME-based method sequentially applies ILRMA and RCSCME to incoming mixture signals.
As described in Section~\ref{ssec:RCSCME}, RCSCME is suitable for real-time processing and can function in real time on modern computers.
However, ILRMA has numerous matrix operations and requires a sufficient number of iterations to converge.
Thus, performing the ILRMA computation within the shift length requires prohibitively expensive computational resources.

To avoid this issue, we focus on the fact that the entire process of the offline RCSCME-based method can be split into the ILRMA and RCSCME parts, and the ILRMA output required in RCSCME is only the demixing matrix $\bm{W}_{i}$, which is a time-invariant parameter that represents the spatial characteristics of the transmission system.
If we can assume that the spatial information does not change abruptly, the estimated $\bm{W}_{i}$ tends to be similar across time frames.
With this assumption, it can be expected that estimating $\bm{W}_i$ every multiple frames does not decrease the separation performance significantly.
Thus, we perform the ILRMA and RCSCME parts in parallel by introducing the blockwise batch algorithm~\cite{Mukai2004IEICE,Mori2006EURASIP}.

Figure~1 shows a schematic of the proposed real-time extension.
In the ILRMA part, the demixing matrix $\bm{W}_{i}$ is estimated from the most recent $\tau_{1}$-s-length observed signals every $\tau_{3}$ s, and the source index containing the target speech in the separated signal is selected on the basis of the maximum kurtosis criterion~\cite{Fujihara2008IWAENC}.
The computational time of these processes can span several frames.
In parallel, the RCSCME part is executed at an interval equal to the shift length, $\tau_{4}$ s, with the most recent $\tau_{2}$-s-length observed signals, the latest $\bm{W}_{i}$, and the estimated target speech index.
Owing to this parallelization, we can avoid the requirement that the ILRMA part must be finished within the shift length, which allows the proposed method to function in real time.

\mysubsection{Incorporation of prior target speech direction information}
\label{ssec:priorinformation}

We have thus far focused on the real-time functioning of the proposed method.
Now, we focus on the channel selection to achieve a reliable separation performance.
As described in Section~\ref{sec:intro}, the kurtosis-based channel selection method often fails and degrades the separation performance in the real-time scenario, although it functions well in the offline scenario.
To resolve this problem, we propose a method utilizing the approximate speaker direction, which is often available in human--avatar/robot communication systems.

In the case where the target speaker talks with the avatar/robot, we can assume that the positional relationship between the target speaker and the microphone array in the avatar/robot (e.g., the target speaker is in front of the avatar/robot) is known in advance.
That is, we can assume that we know in advance the approximate information about the shape of the microphone array and the orientation of the target speaker relative to the microphone array.
To incorporate this information, we replace the usual ILRMA with the ILRMA extended by using the spatial regularization~\cite{Mitsui2018ICASSP}.
It enables us to assign the demixing filter of a specific source index to that for the target speech.
By choosing this source index, we can reduce the channel selection failures and improve the speech extraction performance.
Although it is assumed in \cite{Mitsui2018ICASSP} that steering vectors for all sources are available as prior information, only one steering vector for the target speech is available in our setting.
Thus, we design two regularizers based on only one steering vector.

\mysubsection{Spatial regularization using prior target steering vector}
\label{ssec:SR-ILRMA}

First, we construct the prior $\hat{\bm{A}}_{i}=(\hat{\bm{a}}_{i1},\ldots,\hat{\bm{a}}_{iN})$ corresponding to $\bm{A}_{i}$ from the spatial prior information.
Here, $n^{(\mathrm{t})}$ is defined as the source index of the target speech in the prior information.
That is, the steering vector of the target speech calculated from the prior information is assigned to $\hat{\bm{a}}_{in^{(\mathrm{t})}}$.
To make $\hat{\bm{A}}_{i}$ regular, for example, $\hat{\bm{a}}_{in}\:(n\ne n^{(\mathrm{t})})$ can be set to the basis of the orthogonal complementary space of $\hat{\bm{a}}_{in^{(\mathrm{t})}}$.
Moreover, let $\hat{\bm{W}}_{i}=(\hat{\bm{w}}_{i1},\ldots,\hat{\bm{w}}_{iN})^{\mathsf{H}}\coloneqq\hat{\bm{A}}_{i}^{-1}$ be the prior corresponding to $\bm{W}_{i}$.

In \cite{Mitsui2018ICASSP}, the weighted Euclidean distance between $\bm{W}_{i}$ and $\hat{\bm{W}}_{i}$ is added to the cost function of ILRMA as a regularizer, so that $\bm{W}_{i}$ approaches $\hat{\bm{W}}_{i}$.
However, since only the steering vector of the target speech is available and the method of determining the other components is arbitrary, using $\hat{\bm{W}}_{i}$ for regularization may affect the separation performance.
Then, we first consider using $\hat{\bm{A}}_{i}$ as a supervisor.
Here, we use the fact that $\bm{W}_{i}$ taking around $\hat{\bm{W}}_{i}$ can be regarded as $\bm{W}_{i}\hat{\bm{A}}_{i}$ taking values around the identity matrix $\bm{E}_{N}$.
Therefore, we consider the distances of these matrices, $\| \bm{W}_{i} \hat{\bm{A}}_{i} - \bm{E}_{N} \|^{2}_{\mathrm{F}} = \sum_{n,n'} | \bm{w}_{in}^{\mathsf{H}} \hat{\bm{a}}_{in'} - \delta_{nn'} |^{2}$, and use their weighted ones as regularizers~\cite{Li2020ICASSP}, where $\|\cdot\|_{\mathrm{F}}$ and $\delta_{nn'}$ denote the Frobenius norm and Kronecker's delta, respectively.

Furthermore, to eliminate the effect of $\hat{\bm{a}}_{in}\:(n\ne n^{(\mathrm{t})})$, we remove the regularizers corresponding to $\hat{\bm{a}}_{in}\:(n\ne n^{(\mathrm{t})})$.
We can define the cost function $\mathcal{T}_{\mathrm{SR}}$ of ILRMA with spatial regularization using only $\hat{\bm{a}}_{in^{(\mathsf{t})}}$:
\begin{align}
  \label{costSRstrbased}
  \mathcal{T}_{\mathrm{SR}} = \mathcal{T}_{\mathrm{ILRMA}} +  \sum_{i,n} \mu_{in}^{(\mathrm{SR})} | \bm{w}_{in}^{\mathsf{H}} \hat{\bm{a}}_{in^{(\mathsf{t})}} - \delta_{nn^{(\mathsf{t})}} |^{2},
\end{align}
where $\mu_{in}^{(\mathrm{SR})}$ denotes the weight of the regularizer.

Since the cost function~(\ref{costSRstrbased}) contains a linear term of $\bm{W}_{i}$ in addition to the quadratic form and the logarithm of the determinant, it cannot be optimized by IP as in ILRMA.
In this case, we can apply vectorwise coordinate descent (VCD)~\cite{Mitsui2018ICASSP,Makishima2021SP}, and the update rule for $\bm{w}_{in}$ is as follows:
\begin{align}
  \label{updateDinSR}
  \hat{\bm{D}}_{in}^{(\mathrm{SR})} &\leftarrow \frac{1}{J}\sum_{j}\frac{\bm{x}_{ij}\bm{x}_{ij}^{\mathsf{H}}}{r_{ijn}}+\mu_{in}^{(\mathrm{SR})}\hat{\bm{a}}_{in^{(\mathsf{t})}}\hat{\bm{a}}_{in^{(\mathsf{t})}}^{\mathsf{H}},
  \\
  \label{updateuinSR}
  \bm{u}_{in}^{(\mathrm{SR})} &\leftarrow \bigl( \bm{W}_{i}\hat{\bm{D}}_{in}^{(\mathrm{SR})} \bigr)^{-1}\bm{e}_{n},
  \\
  \label{updateu^inSR}
  \hat{\bm{u}}_{in}^{(\mathrm{SR})} &\leftarrow \mu_{in}^{(\mathrm{SR})} \delta_{nn^{(\mathsf{t})}} \bigl(\hat{\bm{D}}_{in}^{(\mathrm{SR})}\bigr)^{-1} \hat{\bm{a}}_{in},
  \\
  \label{updatehinSR}
  h_{in}^{(\mathrm{SR})} &\leftarrow \bigl(\bm{u}_{in}^{(\mathrm{SR})}\bigr)^{\mathsf{H}} \hat{\bm{D}}_{in}^{(\mathrm{SR})} \bm{u}_{in}^{(\mathrm{SR})},
  \\
  \label{updateh^inSR}
  \hat{h}_{in}^{(\mathrm{SR})} &\leftarrow \bigl(\bm{u}_{in}^{(\mathrm{SR})}\bigr)^{\mathsf{H}} \hat{\bm{D}}_{in}^{(\mathrm{SR})} \hat{\bm{u}}_{in}^{(\mathrm{SR})},
  \\
  \label{updatewinSR}
  \bm{w}_{in} &\leftarrow \begin{cases} 
    \frac{\bm{u}_{in}^{(\mathrm{SR})}}{\sqrt{h_{in}^{(\mathrm{SR})}}} + \hat{\bm{u}}_{in}^{(\mathrm{SR})}, \hspace*{5.5em} (\mathrm{if}\:\hat{h}_{in}=0)\\ 
    \frac{\hat{h}_{in}^{(\mathrm{SR})}}{2h_{in}^{(\mathrm{SR})}}\Biggl[\sqrt{1+\frac{4h_{in}^{(\mathrm{SR})}}{|\hat{h}_{in}^{(\mathrm{SR})}|^{2}}}-1\Biggr]\bm{u}_{in}^{(\mathrm{SR})}\!+\hat{\bm{u}}_{in}^{(\mathrm{SR})}.\\
    \vspace{-0.5em}
    \phantom{\frac{\bm{u}_{in}^{(\mathrm{SR})}}{\sqrt{h_{in}^{(\mathrm{SR})}}} + \hat{\bm{u}}_{in}^{(\mathrm{SR})}} \hspace*{6em} (\mathrm{otherwise})
  \end{cases}
\end{align}

\mysubsection{ILRMA with null-based spatial regularization}
\label{ssec:NSR-ILRMA}
In Section~\ref{ssec:SR-ILRMA}, we cannot use IP for updating $\bm{W}_i$ because \eqref{costSRstrbased} contains the linear term of $\bm{w}_{in}$.
Since VCD is more computationally costly than IP, it may prevent the proposed method described in Section~\ref{ssec:SR-ILRMA} from functioning in real time under limited computational resources.
Thus, we design another regularizer by removing the term that satisfies $\delta_{nn^{(\mathrm{t})}}=1$ in \eqref{costSRstrbased}.
This regularizer can be justified by the following reasons:
in the regularizer of (6), the term that satisfies $\delta_{nn^{(\mathrm{t})}}=1$ is regarded as the regularizer for the scale of the demixing filter for the target speech $\bm{w}_{in^{(\mathrm{t})}}$, whereas the terms that satisfy $\delta_{nn^{(\mathrm{t})}}=0$ induce the demixing filters for noise $\bm{w}_{in}\:(n \ne n^{(\mathrm{t})})$ to form the null of the prior steering vector for the target speech $\hat{\bm{a}}_{in^{(\mathrm{t})}}$.
Here, the term that satisfies $\delta_{nn^{(\mathrm{t})}}=1$ is less important than the other terms because the scale of the demixing filter for the target speech can be modified later by the projection back method~\cite{Murata2001Neurocomputing}.
We call this regularization \textit{the null-based spatial regularization}.
The resultant cost function $\mathcal{T}_{\mathrm{NSR}}$ is defined as
\begin{align}
  \label{costNSR}
  \mathcal{T}_{\mathrm{NSR}} = \mathcal{T}_{\mathrm{ILRMA}} + \sum_{i,n} \mu_{in}^{(\mathrm{NSR})} (1-\delta_{nn^{(\mathsf{t})}}) | \bm{w}_{in}^{\mathsf{H}} \hat{\bm{a}}_{in^{(\mathsf{t})}} |^{2},
\end{align}
where $\mu_{in}^{(\mathrm{NSR})}$ represents the weight of the regularizer.

Since this cost function~(\ref{costNSR}) contains only the quadratic form and the logarithm of the determinant, we can apply the IP to (\ref{costNSR}), and the update rule for $\bm{w}_{in}$ is as follows:
\begin{align}
  \label{updateDinNSR}
  D_{in}^{(\mathrm{NSR})} &\leftarrow \frac{1}{J} \sum_{j} \frac{\bm{x}_{ij} \bm{x}_{ij}^{\mathsf{H}}}{r_{ijn}} \!+ \!\mu_{in}^{(\mathrm{NSR})} (1\!-\!\delta_{nn^{(\mathsf{t})}}) \hat{\bm{a}}_{in^{(\mathsf{t})}} \hat{\bm{a}}_{in^{(\mathsf{t})}}^{\mathsf{H}},
  \\
  \label{updateuinNSR}
  \bm{u}_{in}^{(\mathrm{NSR})} &\leftarrow \bigl(\bm{W}_{i} D_{in}^{(\mathrm{NSR})}\bigr)^{-1} \bm{e}_{n},
  \\
  \label{updatewinNSR}
  \bm{w}_{in} &\leftarrow \frac{\bm{u}_{in}^{(\mathrm{NSR})}}{\sqrt{\bigl(\bm{u}_{in}^{(\mathrm{NSR})}\bigr)^{\mathsf{H}} D_{in}^{(\mathrm{NSR})} \bm{u}_{in}^{(\mathrm{NSR})}}}.
\end{align}

Note that \cite{Goto2022EUSIPCO} has shown that in independent vector analysis with spatial regularization using all $\hat{\bm{a}}_{in}$, the separation performance of the method that removes the terms satisfying $\delta_{nn'}=1$ was equal to or higher than that of the method that uses all the terms.

\mysection{Experiments}

\mysubsection{Experimental conditions}

We evaluate the proposed methods in terms of speech extraction performance and processing time.
Diffuse noise and impulse responses for the target speech were recorded at the Ito International Research Center, The University of Tokyo.
A circular microphone array with a radius of 3.25~cm and equipped with four omnidirectional microphones was placed at a height of 1~m from the floor.
During the diffuse noise recording, 10 participants sat 2--4~m apart from the microphone array and talked to other people around them or read a text given to them beforehand.
At the same time, music was played from loudspeakers embedded in the ceiling.
The impulse responses were recorded under the following conditions:
the height of the target speaker was 1.1~m, the horizontal distance between the target speaker and the microphone array was 1~m, and the reverberation time was $T_{60} = 750$~ms.
We concatenated 50 speech signals from the JSUT dataset~\cite{Takamichi2020AST} with 3-s-length silence intervals in between, creating a total of 226~s of dry source.
Then, the dry source was convolved with the recorded impulse responses and mixed with diffuse noise to simulate mixture signals under diffuse noise.
The signals were mixed so that the input SNR became 0~dB for the entire signals except for the silence interval.
The sampling rate was 16~kHz.
STFT was performed using a 64-ms-length Hann window with a shift length of 32~ms.

We compared three real-time RCSCME-based speech extraction methods using different ILRMAs.
\emph{NaiveILRMA} uses the usual ILRMA (without any spatial prior information) and the kurtosis-based channel selection method.
\emph{SR-} and \emph{NSR-ILRMAs} use the ILRMA extensions proposed in Sections~\ref{ssec:SR-ILRMA} and \ref{ssec:NSR-ILRMA}, respectively, for which we assign the source index $n^{(\mathrm{t})}$ to the target speech.

The RCSCME parts of all the methods are the same.
For each method, the ILRMA part was performed every 512~ms using the most recent 5-s-length signals and the RCSCME part was performed every 32~ms (shift length) using the most recent 3-s-length signals.
The number of NMF bases was set to 10.
For SR- and NSR-ILRMAs, the weight parameters of the regularizers, $\mu_{in}^{(\mathrm{SR})}$ and $\mu_{in}^{(\mathrm{NSR})}$, were both set to 0.1.
For the inverse gamma distribution used in RCSCME, the shape and scale parameters were set to $1.6$ and $10^{-16}$, respectively.
The NMF variables were initialized with uniform random numbers.
The demixing matrix was initialized to the identity matrix for NaiveILRMA and to $\hat{\bm{W}}_{i}$ for SR- and NSR-ILRMAs.
These methods were implemented in Python, and the computation was performed on an Intel Core i9-13900KF CPU, 128~GB RAM, and an NVIDIA GeForce RTX 4090 GPU.
ILRMA and RCSCME computations were performed on the GPU.

The evaluation measures were the processing time and the source-to-distortion ratio~(SDR) \cite{Vincent2006TASLP} improvement.
The groundtruth signal, the observed signal, and the extracted signal were divided by one second each, and the SDR improvement was calculated for each segment.
The intervals whose power of the true target speech signal was less than $10^{-10}$ were excluded, resulting in 184 intervals to be evaluated.
The ILRMA part was executed 442 times for each method to process the entire input signals.
We used the processing time for each execution to evaluate the processing time of the ILRMA part.

\mysubsection{Experimental results}

The processing time of the RCSCME part for all methods was not significantly different.
For all methods together, the average, standard deviation, and maximum of the processing time for the RCSCME part were 7.39, 1.99, and 19.1~ms, respectively.
Since the maximum processing time is less than the shift length of the STFT, all the methods can function in real time.
For NaiveILRMA, whose ILRMA part has the lowest computational cost among all the methods, the average and standard deviation of the processing time for the ILRMA part were 50.9 and 3.85~ms, respectively.
Since the average processing time was longer than the shift length, it was necessary to execute the ILRMA part over multiple frames to achieve real-time operation.

Figure 2 shows a boxplot of processing time for the ILRMA part of each method.
\begin{figure}[t]
  \begin{center}
    \includegraphics[width=0.9\columnwidth]{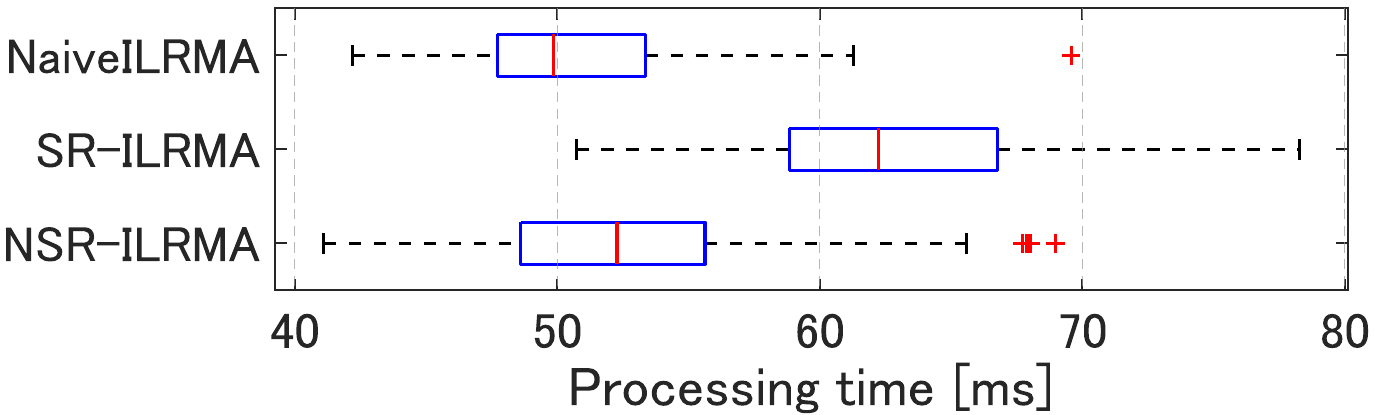}
    \vspace*{-0.2cm}
    \caption{Boxplots of processing time for ILRMA part of all methods.}
    \vspace*{-0.3cm}
  \end{center}
\end{figure}
SR-ILRMA has a longer processing time than NaiveILRMA and NSR-ILRMA.
This is probably because SR-ILRMA uses VCD for parameter updates, whereas NaiveILRMA and NSR-ILRMA use IP.
In addition, NSR-ILRMA is slightly slower than NaiveILRMA.
Although both methods use IP, the update of $D_{in}^{(\mathrm{NSR})}$ in (\ref{updateDinNSR}) is more computationally expensive than that of $D_{in}$ in (\ref{updateDinNaive}).
We consider that this difference in computational complexity caused a slightly longer processing time in NSR-ILRMA.

Next, a boxplot of the SDR improvement for each method is shown in Fig.~3.
\begin{figure}[t]
  \begin{center}
    \includegraphics[width=0.9\columnwidth]{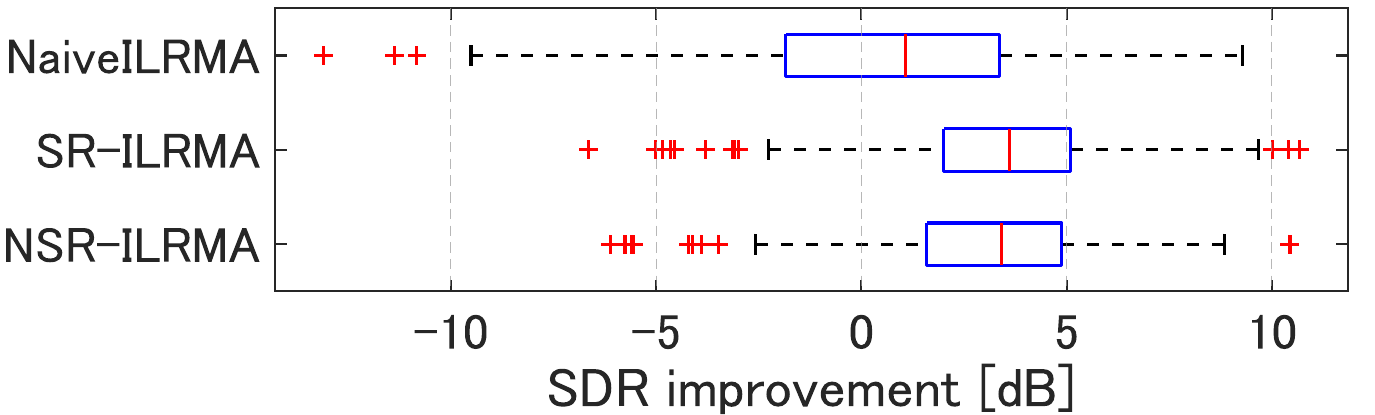}
    \vspace*{-0.2cm}
    \caption{Boxplots of SDR improvement for all methods.}
    \vspace*{-0.2cm}
  \end{center}
\end{figure}
SR- and NSR-ILRMAs outperformed NaiveILRMA and achieved similar SDR improvements.
In addition, there were only a few cases of extremely low SDR improvements in SR- and NSR-ILRMAs, whereas NaiveILRMA often showed such cases.
These results suggest that the channel selection errors did not occur frequently in SR- and NSR-ILRMAs compared with NaiveILRMA, indicating that the use of spatial prior information helps reduce the channel selection errors.
We consider that NSR-ILRMA is more useful than the other methods in terms of both separation performance and computation time.

\mysection{Conclusion}

In this paper, we proposed the real-time RCSCME-based speech extraction method under diffuse noise conditions.
The proposed method was constructed by applying the blockwise batch algorithm to the offline RCSCME-based method.
Furthermore, we incorporated prior information about the approximate target speaker direction into the real-time RCSCME-based method.
On the basis of this information, we designed two regularizers and applied them to the ILRMA extension with spatial regularization.
Speech extraction experiments demonstrated that all the proposed methods function in real time and the designed regularizers improve the speech extraction performance.

\newpage

\bibliographystyle{IEEEbib}
\bibliography{reference}

\end{document}